\documentclass[conference]{IEEEtran}
\IEEEoverridecommandlockouts

\usepackage{cite}
\usepackage{amsmath,amssymb,amsfonts}
\usepackage{algorithmic}
\usepackage{graphicx}
\usepackage{textcomp}
\usepackage{xcolor}
\usepackage[hidelinks]{hyperref}
\usepackage{orcidlink}
\usepackage{subcaption}
\usepackage{braket}
\usepackage{bm}
\usepackage{siunitx}
\usepackage{booktabs}

\def\BibTeX{{\rm B\kern-.05em{\sc i\kern-.025em b}\kern-.08em
    T\kern-.1667em\lower.7ex\hbox{E}\kern-.125emX}}
\begin{document}

\title{Unfair Sampling of Quantum Annealing in Weighted Graph Bipartitioning Problems
    \thanks{
        This work was partially supported by the Japan Society for the Promotion of Science (JSPS) KAKENHI (Grant Number JP23H05447), the Council for Science, Technology, and Innovation (CSTI) through the Cross-ministerial Strategic Innovation Promotion Program (SIP), ``Promoting the application of advanced quantum technology platforms to social issues'' (Funding agency: QST), Japan Science and Technology Agency (JST) (Grant Number JPMJPF2221).
    }
}

\author{
    \IEEEauthorblockN{Shunta Ide \orcidlink{0009-0007-6517-3374}}
    \IEEEauthorblockA{
        \textit{Graduate School of Science and Technology,} \\
        \textit{Keio University} \\
        Kanagawa, Japan
    }
    \and
    \IEEEauthorblockN{Shu Tanaka \orcidlink{0000-0002-0871-3836}}
    \IEEEauthorblockA{
        \textit{Dept. of Appl. Phys. and Physico-Informatics,} \\
        \textit{Graduate School of Science and Technology,} \\
        \textit{KSQAIC,} \\
        \textit{WPI-Bio2Q,} \\
        \textit{Keio University} \\
        Kanagawa, Japan
    }
}

\maketitle

\newcommand{\upa}{\uparrow}
\newcommand{\downa}{\downarrow}

\begin{abstract}
    Quantum annealing (QA) is a promising approach for solving combinatorial optimization problems; however, it is known to exhibit unfair sampling, in which degenerate ground states are not sampled with equal probability even for sufficiently long annealing times.
    Fair sampling is important in applications such as solution diversity assessment and combinatorial counting, yet the mechanism of unfair sampling remains poorly understood, particularly in constrained combinatorial optimization problems.
    In this work, we investigate unfair sampling of QA in weighted graph bipartitioning problems (GBP), a representative constrained optimization problem.
    We study how the penalty coefficient in the penalty method affects sampling fairness.
    Through numerical simulations and experiments on the D-Wave Advantage2 system, we show that increasing the penalty coefficient reduces unfair sampling in a representative single instance, and that this qualitative behavior is also observed on actual hardware.
    A scaling analysis over randomly generated instances with up to 12 spins reveals that, while this trend does not hold universally, more than 70\% of instances exhibit monotonically increasing sampling fairness as the penalty coefficient increases, even at the largest system size studied.
    These results show that increasing the penalty coefficient improves sampling fairness, though at the cost of ground-state probability under practical annealing conditions, and call for a deeper theoretical understanding of unfair sampling in constrained optimization problems.
\end{abstract}


\section{Introduction}
Quantum annealing (QA) is a promising approach for solving combinatorial optimization problems~\cite{kadowaki1998quantum,chakrabarti2022quantum}.
In the standard transverse-field formulation, QA starts from the ground state of the driver Hamiltonian and gradually evolves the Hamiltonian from the driver Hamiltonian to the problem Hamiltonian.
In the ideal closed-system setting, the adiabatic theorem guarantees that, under sufficiently slow evolution, the system remains in the instantaneous ground state throughout the annealing process~\cite{farhi2001quantum}.
Beyond theoretical analysis, commercial QA devices such as the D-Wave system have been applied to a variety of optimization tasks.
In many applications, however, obtaining just one optimal solution is not sufficient.
When the problem has degenerate optima, it is often important to sample all ground states with the same probability, for example in diversity-aware optimization and counting-related tasks~\cite{jerrum1986random,douglass2015constructing,azinovic2017assessment}.

An important issue in this context is that degenerate ground states are generally not sampled uniformly in QA.
This phenomenon, known as unfair sampling, persists even in the adiabatic limit~\cite{matsuda2009quantum,matsuda2009ground,mandra2017exponentially,konz2019uncertain,maruyama2026graph}.
Previous studies have shown that, in geometrically frustrated Ising models, states with more free spins can be favored~\cite{matsuda2009ground}, and that the sampling bias can be modified by using more complex drivers such as higher-order transverse fields~\cite{matsuda2009ground,konz2019uncertain}.
On D-Wave hardware, the required minor embedding has also been reported to affect the degree of unfair sampling~\cite{maruyama2026graph}.
These observations have motivated theoretical analyses based on degenerate perturbation theory, but the mechanism is still not fully understood, especially beyond simplified settings~\cite{konz2019uncertain,maruyama2026graph}.

This gap is particularly important for constrained combinatorial optimization problems.
Many practically relevant problems include hard constraints, and a standard way to handle them in QA is the penalty method, in which a penalty term is added to the objective Hamiltonian~\cite{lucas2014ising,tanaka2017quantum,harwood2021formulating,mugel2022portfolio,glover2019qubo,tanahashi2019application,ide2025extending}.
The penalty coefficient must be chosen to enforce feasibility, but changing this coefficient also alters the low-energy structure explored during annealing.
Despite its practical importance, the influence of the penalty coefficient on unfair sampling in constrained problems has not been systematically investigated.

In this work, we address the following question: how does the penalty coefficient in the penalty method affect the sampling fairness of QA in constrained optimization problems?
We focus on weighted GBP as a representative testbed, since the equal-partition constraint is common to all instances, facilitating analysis of the penalty coefficient.
We first analyze a single instance through numerical simulations and D-Wave Advantage2 experiments to characterize how the penalty coefficient affects sampling fairness.
We then perform a scaling analysis over randomly generated instances with up to 12 spins to examine how broadly this trend holds.

\section{Method}\label{sec:method}
QA in the transverse-field Ising model is an algorithm for finding the ground state of an Ising Hamiltonian.
Many combinatorial optimization problems can be reformulated as finding the ground state of an Ising Hamiltonian~\cite{lucas2014ising}, allowing them to be solved directly by QA.
The objective function is encoded as the problem Hamiltonian:
\begin{equation}
    H_{\mathrm{p}} = -\sum_{i=1}^N h_i \sigma_i^z - \sum_{1\leq i < j \leq N} J_{ij} \sigma_i^z \sigma_j^z,
\end{equation}
where \(N\) is the number of spins, \(\sigma_i^z\) is the Pauli-Z operator of the \(i\)-th spin, \(h_i\) is the magnetic field of the \(i\)-th spin, and \(J_{ij}\) is the coupling strength between the \(i\)-th and \(j\)-th spins.
The total Hamiltonian is typically written as
\begin{equation}
    H(t) = \dfrac{t}{T} H_{\mathrm{p}} + \left(1 - \dfrac{t}{T}\right) H_{\mathrm{q}},
\end{equation}
where \(T\) is the annealing time, and \(H_{\mathrm{q}}\) is the driver Hamiltonian, which is typically given by
\begin{equation}
    H_{\mathrm{q}} = -\sum_{i=1}^N \sigma_i^x,
\end{equation}
where \(\sigma_i^x\) is the Pauli-X operator of the \(i\)-th spin.
During QA, the equal superposition over all computational basis states, which is the ground state of the driver Hamiltonian, is prepared at \(t=0\).
Then, the effect of the driver Hamiltonian is gradually decreased until \(t=T\).
When the annealing time is long enough, the adiabatic theorem ensures that the system stays in the ground state of the instantaneous Hamiltonian during the whole annealing process~\cite{farhi2001quantum}.
Therefore, the ground state of the original optimization problem is expected to be obtained.

For constrained combinatorial optimization problems, constraints must be incorporated into the problem Hamiltonian.
A common approach is the penalty method, in which a penalty term is added to penalize infeasible solutions.
The problem Hamiltonian is then given by
\begin{equation}\label{eq:problem-hamiltonian-mu}
    H_{\mathrm{p}} = H_{\mathrm{obj}} + \mu H_{\mathrm{const}},
\end{equation}
where \(H_{\mathrm{obj}}\) is the original objective function, \(H_{\mathrm{const}}\) is the penalty function, and \(\mu\) is the penalty coefficient.
The penalty function takes a positive value when a spin configuration violates the constraint and 0 when it satisfies the constraint.
The penalty coefficient \(\mu\) is a positive parameter that controls the strength of the penalty.
By adding this penalty term with a sufficiently large \(\mu\), the ground states of the problem Hamiltonian correspond to feasible optimal solutions.
On the other hand, an excessively large \(\mu\) reduces the relative energy differences among feasible solutions and can shrink the minimum energy gap, thereby degrading the performance of QA\@.
To systematically study this trade-off, we decompose \(\mu\) as \(\mu = \mu_{\mathrm{opt}} + \mu_{+}\), where \(\mu_{\mathrm{opt}}\) is the instance-dependent minimum penalty coefficient such that the ground states of \(H_{\mathrm{p}}\) are feasible optimal solutions, and \(\mu_{+} \geq 0\) is the additional penalty beyond this threshold.

We apply the above framework to the (weighted) graph bipartitioning problem (GBP), a representative constrained combinatorial optimization problem.
The GBP seeks a partition of the vertices of an undirected graph into two equal-sized subsets that minimizes the total weight of edges across the cut.
Let \(G = (V, E)\) be an undirected graph with \(N = |V|\) vertices and edge set \(E\), and let \(w_{ij}\) denote the weight of edge \((i, j)\).
Since an equal partition requires exactly \(N/2\) vertices in each subset, a feasible solution exists only when \(N\) is even; we therefore restrict our analysis to even \(N\) throughout this work.
The objective function of the GBP is given by
\begin{equation}
    H_{\mathrm{obj}} = \sum_{(i,j) \in E} w_{ij} \dfrac{1-\sigma_i^z \sigma_j^z}{2}.
\end{equation}
When all edge weights are equal to 1, the problem is called an unweighted GBP; otherwise, it is called a weighted GBP.
The constraint requires that the two subsets contain equal numbers of vertices, which is given by
\begin{equation}
    H_{\mathrm{const}} = \left( \sum_{i \in V} \sigma_i^z \right)^2.
\end{equation}

\section{Results and Discussion}\label{sec:result}
\begin{figure}[!t]
    \centering
    \includegraphics[width=0.8\linewidth]{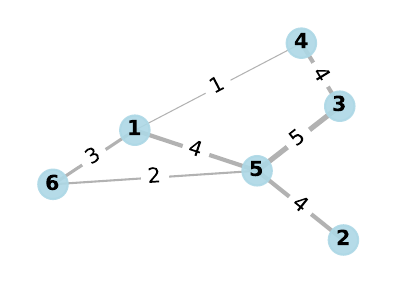}
    \caption{
        An instance of the GBP.
        The number written on the edge \( (i,j) \)  is the weight of the edge \( w_{ij}  \) .
        The optimal solution is to partition the vertices set \((1,2,3,4,5,6)\) into \((2,3,4)\) and \((1,5,6)\), or into \((1,4,6)\) and \((2,3,5)\), with the optimal objective value being 10 in each case.
    }\label{fig:instance}
\end{figure}

We first examined unfair sampling in a single instance shown in Fig.~\ref{fig:instance}.
This instance has four ground states, which reduce to two distinct states when the global spin-flip symmetry is taken into account.
The optimal solution is to partition the vertices into \((2,3,4)\) and \((1,5,6)\), or into \((1,4,6)\) and \((2,3,5)\), with the optimal objective value being 10 in each case.
We solve the time-dependent Schr\"odinger equation using QuTiP~\cite{qutip5,johansson2012qutip,johansson2013qutip}.
To match the energy scale of the D-Wave Advantage2 System 1.13, we normalize \( H_{\mathrm{p}} \) by the auto-scaling factor defined in~\cite{dwave-auto-scale}, excluding the group coupling limit to remove dependence on the embedding of the problem graph onto the hardware topology.

\begin{figure}[!t]
    \centering
    \begin{minipage}{0.9\linewidth}
        \centering
        \includegraphics[width=0.98\linewidth]{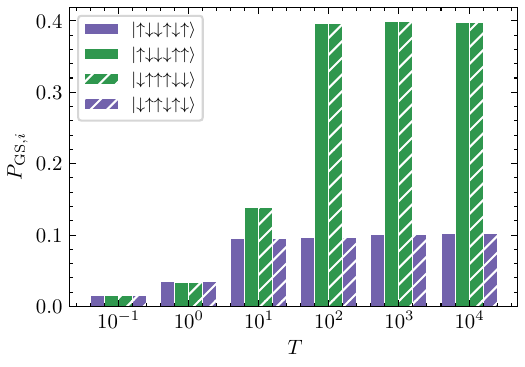}
        \subcaption{}
    \end{minipage}
    \\[2mm]
    \begin{minipage}{0.9\linewidth}
        \centering
        \includegraphics[width=0.98\linewidth]{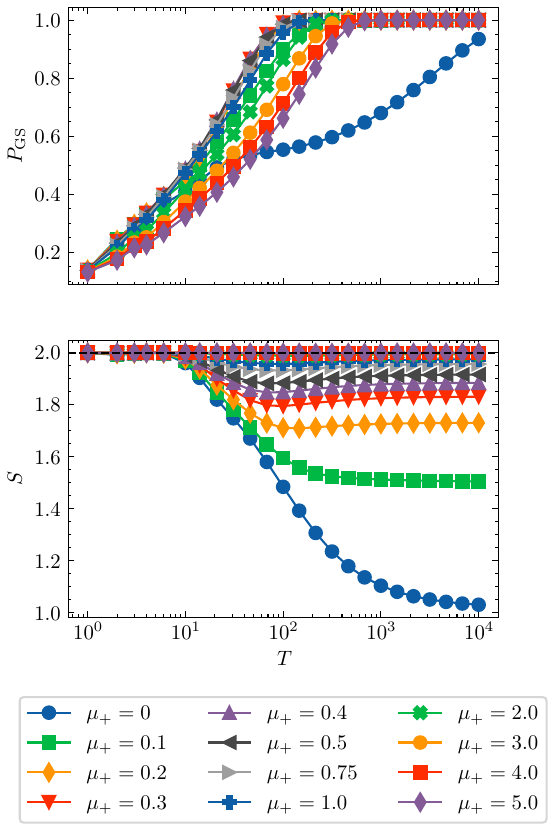}
        \subcaption{}
    \end{minipage}
    \caption{
        The unfair sampling of the GBP instance shown in Fig.~\ref{fig:instance}.
        Panel (a) shows the annealing-time dependence of the ground-state distribution when \(\mu_{+}=0.2\).
        The same color with different hatching pattern represents two states connected by global spin-flip.
        Panel (b) shows the dependence on the penalty coefficient of the probability of obtaining the ground states \(P_{\mathrm{GS}}\) and the (Shannon) entropy \(S\).
    }
    \label{fig:instance-entropy}
\end{figure}

The results are shown in Fig.~\ref{fig:instance-entropy}.
Panel (a) shows the ground-state probability distribution with respect to the annealing time \( T \) when the penalty coefficient is \(\mu_{+}=0.2\), and panel (b) shows the dependence of the ground-state probability \( P_{\mathrm{GS}}  \) and the entropy \( S \) on the penalty coefficient with respect to the annealing time \( T \).
To quantify sampling fairness, we use the Shannon entropy \( S \) of the normalized ground-state probability distribution:
\begin{equation}
    S = -\sum_{i=1}^D p_i \log_2 p_i, \quad p_i = \frac{P_{\mathrm{GS},i}}{P_{\mathrm{GS}}},
\end{equation}
where \(D\) is the number of degenerate ground states (\(D=4\) for this instance), \(P_{\mathrm{GS},i}\) is the probability of obtaining ground state \(i\) that corresponds to an optimal solution, and \(P_{\mathrm{GS}}=\sum_{i=1}^{D} P_{\mathrm{GS},i}\) is the total ground-state probability.
The entropy \(S\) takes its maximum value of \(\log_2 D\) when \(p_i = 1/D\) for all \(i\), corresponding to fair sampling, and decreases as sampling becomes more biased.

Figure~\ref{fig:instance-entropy} (a) shows that the ground-state probability distribution is not uniform even when the annealing time is long enough for \(P_{\mathrm{GS}}\) to reach unity, for \(\mu_{+} = 0.2\).
The two ground states related by the global spin-flip symmetry (\(\ket{\upa\downa\downa\downa\upa\upa}\) and \(\ket{\downa\upa\upa\upa\downa\downa}\)) each converges to a probability of approximately 0.4, while the remaining pair (\(\ket{\upa\downa\downa\upa\downa\upa}\) and \(\ket{\downa\upa\upa\downa\upa\downa}\)) each converges to approximately 0.1.
Figure~\ref{fig:instance-entropy} (b) shows the dependence of the ground-state probability \(P_{\mathrm{GS}}\) and the entropy \(S\) on the penalty coefficient \(\mu_{+}\) as functions of the annealing time \(T\).
For all \(\mu_{+} > 0\), \(P_{\mathrm{GS}}\) reaches unity by \(T \approx 10^{2}\)--\(10^{3}\), whereas for \(\mu_{+} = 0\), \(P_{\mathrm{GS}}\) remains below 1 even at \(T = 10^{4}\), which is attributed to the smaller energy gap between the ground state and the first excited state (the lowest infeasible state).
For \(\mu_{+} = 0\), \(S\) decreases monotonically from \(\log_2 D = 2\) toward approximately 1.0 as \(T\) increases, indicating strong sampling bias.
As \(\mu_{+}\) increases, this decrease in \(S\) is suppressed and the distribution approaches fair sampling.
These results show that, for short annealing times (\(T \lesssim 10^{3}\)), increasing the penalty coefficient improves sampling fairness at the cost of a reduced ground-state probability, indicating a trade-off between sampling fairness and ground-state probability.
In contrast, in the near-adiabatic regime (\(T \approx 10^{4}\)), \(P_{\mathrm{GS}}\) remains near unity across all \(\mu_{+}\), and increasing the penalty coefficient enhances the entropy without sacrificing ground-state probability, thereby eliminating this trade-off.

Next, we executed QA on D-Wave Advantage2 System 1.13 with the same instance as in the previous result.
We employed parallel annealing, a technique that enables simultaneous annealing processes of the same problem using multiple embeddings of the problem graph onto the hardware-specific graph.
The maximum number of embeddings is 444, and we set it to 400 to avoid long chain lengths in the embeddings.
We also employed the spin reversal transformation (SRT), a (local) gauge transformation technique in which subsets of spins are flipped along with the corresponding couplings and biases, to mitigate systematic errors in couplings and local fields on the hardware and thereby increase the ground-state probability.
The annealing time was set to \SI{200}{\micro\second}, the number of SRT to 10, and the number of reads to 1000.
The annealing time is ten times longer than the default setting to increase the probability of obtaining the ground state.
Therefore, each call returns \(400\times10\times1000=4\times10^6\) samples.
Note that the previous study~\cite{maruyama2026graph} suggested that minor embedding can affect the characteristics of unfair sampling.
In this work, we investigated the sampling behavior by explicitly simulating the embedded problem (including SRT) using QuTiP, restricted to cases where the embedded problem size remains computationally tractable.
In the smallest case, the embedded problem size was 8 spins, and almost all embedded problems remained computationally tractable.
We confirmed that the resulting sampling behavior is qualitatively consistent with that shown in Fig.~\ref{fig:instance-entropy}.

\begin{figure}[!t]
    \centering
    \begin{minipage}{0.98\linewidth}
        \includegraphics[width=0.98\linewidth]{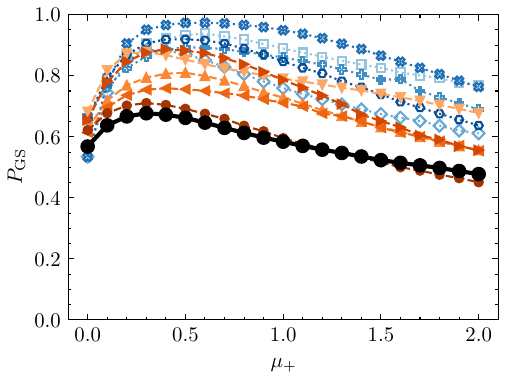}
        \subcaption{}
    \end{minipage}
    \\[4mm]
    \begin{minipage}{0.98\linewidth}
        \includegraphics[width=0.98\linewidth]{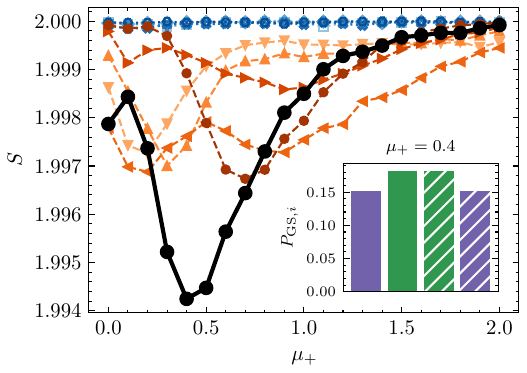}
        \subcaption{}
    \end{minipage}
    \caption{
        The unfair sampling of the GBP instances on the D-Wave system.
        Panels (a) and (b) show the dependence on the penalty coefficient of the ground-state probability \( P_{\mathrm{GS}} \) and the entropy \( S \), respectively.
        The black line is the result for the D-Wave system with the same instance shown in Fig.~\ref{fig:instance}.
        The orange dashed lines are the results for 5 instances which exhibit unfair sampling in the simulation.
        The blue dotted lines are the results for 5 instances which exhibit fair sampling in the simulation.
        The inset in panel (b) shows the ground-state probability distribution at \(\mu_+ = 0.4\), with the legend shared with Fig.~\ref{fig:instance-entropy}~(a).
    }
    \label{fig:dwave-entropy}
\end{figure}

The results are shown in Fig.~\ref{fig:dwave-entropy}.
As a baseline, we present results for 5 instances with \(N=6\) in which all ground states are obtained with equal probability for all penalty coefficients and annealing times in the simulation, and 5 instances that exhibit unfair sampling comparable to that shown in Fig.~\ref{fig:instance-entropy} (b) in the simulation, where the instances are generated by the same procedure as described in the scaling analysis later.
As shown in Fig.~\ref{fig:dwave-entropy} (a), the ground-state probability first increased and then decreased as the penalty coefficient was raised from the optimal value.
The initial increase is considered to be due to the widening of the energy gap between the ground states and the first excited states, which improves the ability of QA to resolve the ground states.
The subsequent decrease is considered to be due to the dominance of the constraint term over the objective function.
As \(\mu_{+}\) grows, constraint satisfaction is prioritized, which suppresses objective minimization and reduces the ground-state probability.
In Fig.~\ref{fig:dwave-entropy} (b), the variation in entropy is much smaller than in the simulation; nevertheless, unfair sampling is still observed.
The entropy first increases slightly from \(\mu_{+}=0\), then decreases to a minimum at \(\mu_{+}=0.4\), and increases again with further increase of the penalty coefficient.
At \(\mu_{+}=0.4\), the probability of obtaining each ground state was \(0.153\), \(0.183\), \(0.182\), and \(0.153\), ordered from top to bottom in the legend of Fig.~\ref{fig:instance-entropy} (a).
The states that are more likely to appear correspond to those identified in the simulation, and the two states with equal probabilities also match those in the simulation.
Compared with the baseline instances that exhibit fair sampling in simulation, the entropy of all such instances is almost equal to 2 for any \(\mu_{+}\) and \(P_{\mathrm{GS}}\); however, the instances that exhibit unfair sampling in simulation show unfair sampling.
Although the degree of unfair sampling is smaller than in the simulation, likely due to noise and thermal relaxation at the finite temperature of the hardware, unfair sampling is still observed in the actual hardware.
As the penalty coefficient is increased, the ground-state probability decreases, while the sampling becomes closer to fair sampling.

We finally perform a scaling analysis over randomly generated instances in simulation.
We generate the problem instances by the following procedure.
For a given number of vertices \( N \), each pair of vertices is connected independently with probability 0.5, and the weight of each edge is generated from a discrete integer uniform distribution between 1 and \( N \).
When the resulting graph is connected, we verify that the ground states are degenerate and the degeneracy equals a desired value.
By repeating the above procedure, we generate 100 problem instances for each \( N \in \{4,6,8,10,12\} \), each having exactly 2 degenerate ground states up to the global spin-flip symmetry.
The simulation is performed by solving the time-dependent Schr\"odinger equation with QuTiP~\cite{qutip5,johansson2012qutip,johansson2013qutip}.
Here, we employed the problem Hamiltonian given by
\begin{equation}\label{eq:problem-hamiltonian-lambda}
    H_{\mathrm{p}} = (1-\lambda)H_{\mathrm{obj}} + \lambda H_{\mathrm{const}}
\end{equation}
where \(\lambda \in [0, 1]\) is a normalized penalty coefficient.
This parameterization is equivalent to Eq.~\eqref{eq:problem-hamiltonian-mu} with \( \mu=\lambda/(1-\lambda) \), up to an overall energy scale factor of \(1/(1-\lambda)\), and covers the full range \( \mu \in [0, \infty) \) while keeping the energy scale bounded.
Specifically, \( \lambda = 0 \) corresponds to \( \mu = 0 \) and \( \lambda = 1 \) corresponds to \( \mu \to \infty \).

\begin{figure}[!t]
    \centering
    \includegraphics[width=0.98\linewidth]{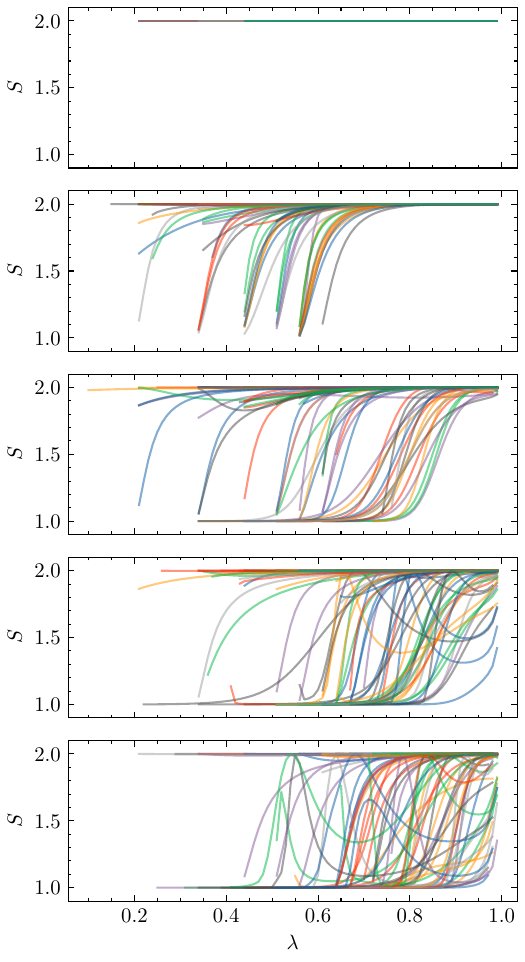}
    \caption{
        The entropy of the ground-state probability distribution with respect to the penalty coefficient \(\lambda\) for 100 randomly generated instances for each number of spins.
        The horizontal axis represents the penalty coefficient \(\lambda\), and the vertical axis shows the entropy of the normalized ground-state probability, for \(N=4, 6, 8, 10, 12\) from top to bottom.
        Note that the entropy is equal to 2 in all instances when \(N=4\).
    }
    \label{fig:multiple-instance-entropy}
\end{figure}

The results are shown in Fig.~\ref{fig:multiple-instance-entropy}.
The horizontal axis represents the penalty coefficient \(\lambda\) ranging from 0 to 1, and the vertical axis shows the entropy of the normalized ground-state probability, for \(N=4, 6, 8, 10, 12\) from top to bottom, respectively.
The entropy is only shown when the ground-state probability is effectively equal to 1.
The annealing time is set to \( T=10^5 \), which is expected to be long enough to reach the ground state with probability 1.
When \( N = 4 \), the entropy is equal to 2, which indicates fair sampling in all instances.
When \( N = 6 \), the entropy increases monotonically as the penalty coefficient is increased.
As \( N \) is increased, the behavior of the entropy becomes more complex.
In some instances, we observe that the entropy decreases once and then increases afterward.

The monotonic increase rate of the entropy with respect to the penalty coefficient \(\lambda\) in each \( N \) is shown in Table~\ref{tab:monotonic-increase-rate}.
As \(N\) increases, the monotonic increase rate decreases and appears to saturate around 70\%--75\% for the system sizes studied, suggesting that increasing the penalty coefficient improves sampling fairness in the majority of cases.
\begin{table}[!t]
    \centering
    \caption{
        The monotonic increase rate of the entropy with respect to the penalty coefficient \(\lambda\) for the multiple-instance results.
        The monotonic increase rate is the fraction of 100 instances in which \(S\) is non-decreasing across all sampled values of \(\lambda\).
    }
    \label{tab:monotonic-increase-rate}
    \begin{tabular}{cc}
        \toprule
        \(N\) & monotonic increase rate \\
        \midrule
        4     & 1.00                    \\
        6     & 0.91                    \\
        8     & 0.74                    \\
        10    & 0.72                    \\
        12    & 0.74                    \\
        \bottomrule
    \end{tabular}
\end{table}

\section{Conclusion}\label{sec:conclusion}
In this study, we investigated the unfair sampling of quantum annealing in weighted graph bipartitioning problems and examined the effect of the penalty coefficient on the sampling behavior.
We confirmed that unfair sampling occurs in QA applied to GBP, and that in a single instance, increasing the penalty coefficient suppresses unfair sampling.
We also verified experimentally on the D-Wave Advantage2 system that unfair sampling persists in actual hardware, and a qualitatively similar dependence on the penalty coefficient is observed there.
A scaling analysis over 100 randomly generated instances for each \(N = 4, 6, 8, 10, 12\) showed that the monotonic increase rate decreases with system size and appears to saturate around 70\%--75\% for the system sizes studied, indicating that the single-instance trend does not generalize universally but that increasing the penalty coefficient improves sampling fairness in the majority of cases.
These findings suggest that the penalty coefficient, traditionally tuned solely for feasibility, also serves as a control parameter for sampling fairness, opening a new perspective on penalty design in constrained quantum optimization.

Several directions remain for future work.
The most pressing open question is the mechanism behind the dependence of unfair sampling on the penalty coefficient; degenerate perturbation theory is a promising framework for this analysis and warrants further investigation in the constrained setting.
The effects of noise and thermal relaxation on unfair sampling in actual hardware also remain to be characterized quantitatively.
It also remains to clarify how unfair sampling changes when the ground-state degeneracy becomes larger, beyond the fourfold-degenerate cases considered in the present study.
Extending the analysis beyond GBP to other constrained combinatorial optimization problems is an important next step.
Finally, structure-aware drivers such as the XY-mixer, which inherently preserves the equal-partition constraint, may alter the unfair sampling behavior.
Although realizing such mixers in quantum annealing hardware is challenging, the quantum alternating operator ansatz with gate-based quantum computers provides a natural framework for their implementation, and investigating unfair sampling under these constraint-preserving drivers is a promising direction.

\section*{Acknowledgment}
S. Tanaka wishes to express their gratitude to the World Premier International Research Center Initiative (WPI), MEXT, Japan, for their support of the Human Biology-Microbiome-Quantum Research Center (Bio2Q).

\bibliographystyle{IEEEtran}
\bibliography{reference}

\end{document}